\documentclass[fleqn,10pt,onecolumn]{wlscirep}
\usepackage{amsfonts}
\usepackage{txfonts}
\usepackage{amsmath}
\usepackage{float}
\usepackage{graphicx}
\usepackage{dcolumn}
\usepackage{bm}
\usepackage{amssymb}
\usepackage{mathrsfs}
\usepackage{amsmath}
\usepackage{bm}
\usepackage{subfigure}
\usepackage{cases}

\title{Quantum synchronization of two mechanical oscillators in coupled optomechanical systems with Kerr nonlinearity}
\author[1,2]{Guo-jian Qiao}
\author[1,2]{Hui-xia Gao}
\author[1,2,*]{Hao-di Liu}
\author[1]{X. X. Yi}
\affil[1]{Center for Quantum Sciences and School of Physics, Northeast Normal University, Changchun 130024, China}
\affil[2]{National Demonstration Center for Experimental Physics Education, Northeast Normal University, Changchun 130024, China}
\affil[*]{Correspondence and requests for materials
should be addressed to H.-D.L. (email: liuhd100@nenu.edu.cn)}

\begin{abstract}
We investigate the quantum synchronization phenomena of two mechanical oscillators of different
frequencies in two optomechanical systems under periodically modulating cavity detunings or driving amplitudes, which can interact mutually through an optical fiber or a phonon tunneling.  The cavities are filled with Kerr-type nonlinear medium. It is found that, no matter which the coupling and  periodically modulation we choose, both of the quantum synchronization of nonlinear optomechanical system are more appealing than the linear optomechanical system. It is easier to observe greatly enhanced quantum synchronization with Kerr nonlinearity. In addition, the different influences on the quantum synchronization between the two coupling ways and the two modulating ways  are compared and discussed.
\end{abstract}
\begin{document}
\flushbottom
\maketitle
\thispagestyle{empty}

\section*{Introduction}

Spontaneous synchronization is one of the most ordinary and valuable phenomena in classical physics, which was firstly noticed by Huygens in the experiments of the oscillations of two pendulum clocks with a common support \cite{huygens1897oeuvres}. In the last decade, synchronization has been widely applied in various fields, e.g., neuron networks \cite{PhysRevE.91.032813,PhysRevE.84.046207,Cao2017}, chemical reactions \cite{vaidyanathan2015dynamics}, heart cells \cite{romashko1998subcellular}, fireflies \cite{pikovsky2003synchronization}, hyperbolic systems\cite{Li2016}. The reason for spontaneous synchronization effect drawing much attention recently is the searching for similar phenomena in quantum regimes. Mari et al. proposed a concept of complete synchronization and phase synchronization for quantum system and gave an effective synchronization measurement scheme in the continuous variable (CV) system \cite{PhysRevLett.111.103605}. Subsequently, this work attracted extensive attention in many physical systems of quantum synchronization, such as optomechanics \cite{PhysRevLett.111.073603,weiss2016noise}, cavity quantum electrodynamics \cite{PhysRevB.80.014519,PhysRevA.91.012301}, atomic ensembles \cite{PhysRevLett.113.154101,PhysRevLett.114.103601,PhysRevA.91.061401}, Van der Pol (VdP) oscillators \cite{PhysRevA.91.012301,PhysRevLett.111.234101,PhysRevE.89.022913,PhysRevLett.112.094102,walter2015quantum} , Bose-Einstein condensation \cite{samoylova2015synchronization}, superconducting circuit systems \cite{gul2014synchronization,PhysRevLett.111.073602}. Moreover, relevant experiments verified the theoretical predictions successfully and a lot of new researches based on application have emerged recently \cite{PhysRevLett.109.233906,PhysRevLett.111.213902,PhysRevLett.112.014101,PhysRevLett.115.163902}.

In Mari's work,  two coupled photomechanical devices was chosen to study the quantum synchronization, since linear optomechanics which explores the coupling between photons and phonons via radiation pressure, have made great progress recently. To realize perfect quantum synchronization in optomechanical system, the existing researches mainly focus on the different ways of coupling between two subsystems: the two mechanical oscillators directly coupled by phonons \cite{PhysRevLett.111.103605,PhysRevE.95.022204} or the two cavity modes coupled through an optical fiber \cite{PhysRevE.95.022204,PhysRevE.93.062221}. Cavity mode and external field can also be modulated by periodic function to achieve better quantum synchronization \cite{du2017synchronization,geng2018enhancement,PhysRevA.80.052316}. But the form of systematic Hamiltonian and cavity mode is unchanged essentially, only through different ways of coupling, as well as to the coupling effect of periodic modulation to implement the energy transmission between the subsystems \cite{PhysRevE.95.022204,geng2018enhancement}. However, the nonlinearity of the optomechanical interaction of the quantum level is also important. In an optomechanical system, nonlinear interaction such as parametric amplifications and optical Kerr effect and nonlinear optical effects in materials are widely concerned \cite{boyd1992nonlinear,zhou2017second,zielinska2017self}. Meanwhile, high-order optomechanically induced transparency effects is also proposed on account of the intrinsic nonlinear optomechanical interactions, such as photon-phonon polariton pairs and sideband generations \cite{fan2015cascaded,PhysRevLett.111.053602,PhysRevLett.111.133601,PhysRevLett.111.083601,PhysRevA.86.013815,PhysRevA.97.013843}. Recent works studied the physics of the nonlinear interaction in weakly driven systems in theory and we can realize quantum nonlinearity into optomechanical systems by the method of a nonlinear optical medium or a nonlinear mechanical oscillator experimentally \cite{aspelmeyer2014cavity}. Hence, one will naturally ask, are the behaviors of the quantum synchronization the same in linear and nonlinear optomechanical system? Does the Kerr nonlinearity can be used as a resource for perfect quantum synchronization?

To shed light on these questions, in this work we study the quantum synchronization phenomenon of two mechanical oscillators of different frequencies in two optomechanical systems with the cavities filled by Kerr-type medium.  The coupling between the two subsystems can be either directly a phonon tunneling or indirectly an optical fiber. The cavity detunings and the driving amplitudes can be alternatively periodically modulated. The enhancement of Kerr nonlinearity to the quantum synchronization are investigated in both the two coupling ways and the two modulation ways. In addition, we also compare and discuss the different effects on the quantum synchronization between the two different coupling ways (indirectly coupled mechanical oscillators through an optical fiber and directly coupled mechanical oscillators by phonon tunneling) and the two modulating ways (periodical modulation on cavity detunings and driving amplitudes)

\section*{Model and Main equations}
\begin{figure}[!hbt]
\centering
\includegraphics*[width=0.5\columnwidth]{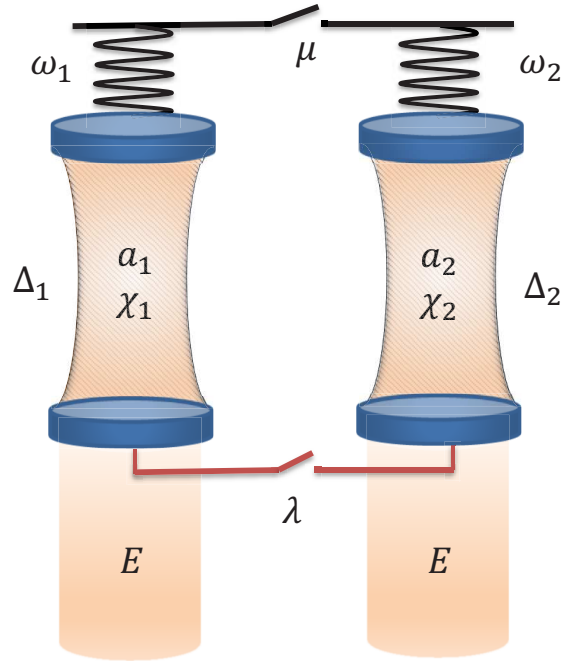}
\caption{ Schematic illustration of two coupled nonlinear optomechanical system. The switches denote that the coupling between the two systems is either the two mechanical oscillators interacting mutually through a phonon tunneling or the two cavity modes coupled through an optical fiber.}
\label{1}
\end{figure}

The system we choose to study the quantum synchronization is modeled by two coupled optomechanical devices. Each optomechanical device consists of a mechanical oscillator coupled with a Fabry-P\'{e}rot cavity filled with Kerr-type nonlinear medium (see Fig.\ref{1}) and driven by a time-periodic modulated filed. The coupling between the two devices can be realized by the interaction of the two mechanical oscillator through a phonon tunneling term of intensity $\mu$ \cite{PhysRevLett.111.073603} or the coupling between the two cavity mode through an optical fiber. Then the Hamiltonian of the whole system takes the form $(\hbar=1)$
\begin{equation}
\begin{aligned}
H=&\sum_{j=1}^{2}\left\{-\triangle_{j}\left[1+\eta_{C}\cos(\Omega_{C} t)\right]a_{j}^{\dagger}a_{j}-\chi_{j}\left(a_{j}^{\dagger}a_{j}\right)^{2}+\frac{\omega_{j}}{2}\left(p_{j}^{2}+q_{j}^{2}\right)-ga_{j}^{\dagger}a_{j}q_{j}\right.\\
   &\left.+iE\left[1+\eta_{D}\cos(\Omega_{D}t)\right]\left(a_{j}^{\dagger}-a_{j}\right)\right\}-\mu q_{1}q_{2}+\lambda\left(a_{1}^{\dagger}a_{2}+ a_{2}^{\dagger}a_{1}\right).
\end{aligned}
\end{equation}
In this expression, $a_{j}$ and $a_{j}^{\dagger}$ are the creation and annihilation
operators for the optical field, $q_{j}$ and $p_{j}$ are
dimensionless position and momentum operators of the $j$-th mechanical oscillator respectively\cite{wang2016steady,jin2017reconfigurable}. $\omega_{j}$ are the mechanical frequencies,  $\triangle_{j}$ are the optical detunings which can be modulated with a common frequency $\Omega_{C}$ and amplitude $\eta_{C}$. $\chi_{j}$ are Kerr coupling
coefficients, $g$ is the optomechanical coupling constant. The driving fields with intensity $E$ are modulated with frequency $\Omega_{D}$ and amplitude $\eta_{D}$. The coupling between the two optomechnical system is chosen as either the interaction between the two mechanical oscillators through a phonon tunneling term of intensity $\mu$, or the cavity modes coupled through an optical fiber with strength $\lambda$. Considering the dissipation of the system, the quantum Langevin equations in the Heisenberg picture of our model can be derived as \cite{farace2012enhancing,genes2009quantum,bai2017classical,du2017synchronization}
\begin{equation}
\begin{aligned}
\dot{q_{j}}& =\omega_{j}p_{j}, \\
 \dot{p_{j}}& =-\omega_{j}q_{j}-
 \gamma p_{j}+ga_{j}^{\dagger}a_{j}+ \mu q_{3-j}+\xi_{j}, \\
 \dot{a_{j}}&=-\left\{\kappa-i\triangle_{j}\left[1+\eta_{C}\cos(\Omega_{C}t)\right]-i\chi_{j}\left(a_{j}^{\dagger}a_{j}+a_{j}a_{j}^{\dagger}\right)\right\}a_{j}+iga_{j}q_{j}+E[1+\eta_{D}\cos(\Omega_{D}t)]-i\lambda a_{3-j}+\sqrt{2\kappa}a_{j}^{in},
\end{aligned}
\label{LE}
\end{equation}
where $\kappa$ and $\gamma$ are the optical and mechanical damping rates, respectively ($\kappa$ and $\gamma$ are assumed to be equal in both systems for simplicity). $a^{in}$ is
the radiation vacuum input noise obeying standard correlation relations $\left\langle a_{j}^{in\dagger}(t)a_{j'}^{in}(t')+a_{j'}^{in}(t')a_{j}^{in\dagger}(t)\right\rangle=\delta_{jj'}\delta(t-t')$. The Brownian noise operator $\xi_{j}$ describes
the stochastic noise operator of one mechanical oscillator satisfying $\frac{1}{2}\langle \xi_{j}(t)\xi_{j'}(t')+\xi_{j'}(t')\xi_{j}(t)\rangle=\gamma(2n_{b}+1)\delta_{jj'}\delta(t-t')$, where $n_{b}=1/\exp\left(\hbar \omega_{j}/k_{B}T-1\right)$ is the mean phonon number of the mechanical bath which gauges the temperature $T$ of the system \cite{giovannetti2001phase,liu2014optimal,xu2015optical}.

To solve the ``classical" (mean values) and ``quantum" parts (fluctuations) of Eq.(\ref{LE}) separately, we adopt the mean-field approximation \cite{li2015criterion,li2016quantum,mari2012opto,mari2009gently} by decomposing every operator as its average value plus a small fluctuation, i.e.
\begin{equation}
  a_j(t)=\alpha_j(t)+\delta a(t), ~~O_j(t)=\bar{O}_j(t)+\delta O_j.~(O=q,p)
\label{MF}
\end{equation}
Substituting (\ref{MF}) into Eq. (\ref{LE}),  we obtain the following ``classical" equations for average values
\begin{equation}
\begin{aligned}
\dot{\bar{q}}_{j}& =\omega_{j}\bar{p}_{j}, \\
\dot{\bar{p}}_{j}& =-\omega_{j}\bar{q}_{j}-\gamma \bar{p}_{j}+g|\alpha_{j}|^{2}+ \mu \bar{q}_{3-j}, \\
 \dot{\alpha_{j}}&=-\left\{\kappa-i\triangle_{j}\left[1+\eta_{C}\cos(\Omega_{C}t)\right]-2i\chi_{j}|\alpha_{j}|^{2}\right\}\alpha_{j}+ig\alpha_{j}\bar{q}_{j}+E[1+\eta_{D}\cos(\Omega_{D}t)]-i\lambda \alpha_{3-j},
\label{mb}
\end{aligned}
\end{equation}
and the ``quantum" equations for fluctuations
\begin{equation}
\begin{aligned}
\dot{\delta q_{j}}& =\omega_{j}\delta p_{j}, \\
\dot{\delta p_{j}}& =-\omega_{j}\delta q_{j}-\gamma\delta p_{j}+g(\alpha_{j}\delta a_{j}^{\dagger}+\alpha_{j}^{*}\delta a_{j})+ \mu \delta q_{3-j} +\xi_{j},\\
\dot{\delta a_{j}}&=-\left\{\kappa-i\triangle_{j}\left[1+\eta_{C}\cos(\Omega_{C}t)\right]\right\}\delta a_{j}+2i\chi_{j}\left\{2|\alpha_{j}|^{2}\delta a_{j}+\alpha^2_{j}\delta a_{j}^{\dagger}\right\}+ig(\alpha_{j}\delta q_{j}+\bar{q}_{j}\delta a_{j})-i\lambda \delta a_{3-j}+\sqrt{2\kappa}a_{j}^{in}.
\end{aligned}
\label{flu}
\end{equation}
where we've ignored the second and the higher order small terms. Taking the transformations of optical field operators $\delta x_{j}=\frac{1}{\sqrt{2}}\left(\delta a_{j}^{\dagger}+\delta a_{j}\right)$, $\delta y_{j}=\frac{1}{\sqrt{2}i}\left(\delta a_{j}^{\dagger}-\delta a_{j}\right)$ and the noise $\delta x_{j}^{in}=\frac{1}{\sqrt{2}}\left(\delta a_{j}^{in^{\dagger}}+\delta a_{j}^{in}\right)$,  $\delta y_{j}^{in}=\frac{1}{\sqrt{2}i}\left(\delta a_{j}^{in^{\dagger}}-\delta a_{j}^{in}\right)$,  Eq.(\ref{flu}) can be writen as
\begin{equation}
  \dot{u}=Mu+n,
  \label{mc}
\end{equation}
with the fluctuation vector $u=(\delta q_{1}, \delta p_{1}, \delta x_{1}, \delta y_{1}, \delta q_{2}, \delta p_{2}, \delta x_{2}, \delta y_{2})^{\top} $, the noise vector $n=(0, \xi_{1}, \kappa, \kappa, 0, \xi_{2}, \kappa, \kappa)^{\top} $ and the time-dependent matrix
\begin{equation}
M=\left(\begin{array}{cccccccc}
0&\omega_{1} & 0&0 & 0 & 0 & 0&0\\
-\omega_{1} &-\gamma&\sqrt{2}g \mathrm{Re}(\alpha_{1})&\sqrt{2}g \mathrm{Im}(\alpha_{1})&\mu &0&0&0\\
\sqrt{2}g\mathrm{ Im}(\alpha_{1})&0&G_1^-&-F_{1}^-&0&0&0&\lambda\\
\sqrt{2}g \mathrm{Re}(\alpha_{1})&0&F^+_{1}&G_1^+ &0&0&-\lambda&0\\
0 & 0 & 0&0& 0 & \omega_{2} & 0&0\\
\mu &0&0&0&-\omega_{2} &-\gamma&\sqrt{2}g \mathrm{Re}(\alpha_{2})&\sqrt{2}g \mathrm{Im}(\alpha_{2})\\
0&0&0&\lambda&\sqrt{2}g\mathrm{ Im}(\alpha_{2})&0&G^-_2&-F_2^-\\
0&0&-\lambda&0& \sqrt{2}g \mathrm{Re}(\alpha_{2})&0&F^+_{2}&G^+_2
\end{array}\right)
\end{equation}
with
\begin{equation}
\begin{aligned}
F^{\pm}_{1,2}&=\triangle_{1,2}[1+\eta_{C}\cos(\Omega_{C}t)]+g\bar{q}_{1,2}\pm\left\{2\chi_{1,2}\left[\mathrm{Re}^{2}(\alpha_{1,2})-\mathrm{Im}^{2}(\alpha_{1,2})\right]+4\chi_{1,2}|\alpha_{1,2}|^{2}\right\},\\
G^{\pm}_{1,2}&=-\kappa\pm4\chi_{1,2}\mathrm{Re}(\alpha_{1,2})\mathrm{Im}(\alpha_{1,2}).
\end{aligned}
\end{equation}
As proposed by Mari et al \cite{PhysRevLett.111.103605}, through a figure of merit
\begin{equation}
 S_{qm}\equiv\left\langle q_{-}^{2}(t)+p_{-}^{2}(t)\right\rangle^{-1}
\label{Mmatrix}
\end{equation}
with the synchronization errors $q_{-}(t)\equiv\frac{1}{\sqrt{2}}[q_{1}(t)-q_{2}(t)]$ and $p_{-}(t)\equiv\frac{1}{\sqrt{2}}[p_{1}(t)-p_{2}(t)]$, the synchronization level of the two mechanical oscillators in the two optomechanical system can be gauged. The Heisenberg principle set the value of $S_q$ ranging from $0$ to $1$ (complete synchronization) \cite{PhysRevLett.111.103605}. With the mean-field treatment above, this generalized synchronization can be extended from the classical to the quantum regime
by excluding the mean value of the conjugate quantities simultaneously, i.e. taking the changes of variables:
\begin{equation}
q_{-}(t) \rightarrow q_{-}(t)-\bar{q}_{-}(t)=\delta q_-(t), ~~  p_{-}(t) \rightarrow p_{-}(t)-\bar{p}_{-}(t)=\delta p_-(t).
\end{equation}
Therefore, the mean values of quantum errors $\langle \delta q_{1}-\delta q_{2}\rangle$ and $\langle \delta p_{1}-\delta p_{2}\rangle$ arising from the noise terms can be used to measure the quantum synchronization as \cite{li2016quantum,du2017synchronization}
\begin{equation}
\begin{aligned}
S_{q}(t)&=\left\langle \delta q_{-}^{2}(t)+\delta p_{-}^{2}(t)\right\rangle^{-1}\\
&=\left\langle  \delta q_{1}^{2}(t)+\delta q_{2}^{2}(t)+\delta p_{1}^{2}(t)+\delta p_{2}^{2}(t)-\delta q_{1}(t)\delta q_{2}(t)-\delta q_{2}(t)\delta q_{1}(t)-\delta p_{1}(t)\delta p_{2}(t)-\delta p_{2}(t)\delta p_{1}(t)        \right\rangle^{-1}\\
\end{aligned}
\end{equation}
if we define the mean values of the quantum fluctuations by a $8 \times 8$
covariance matrix
\begin{equation}
V_{ij}\equiv\frac{1}{2}\langle u_{i}u_{j}+u_{j}u_{i}\rangle
\end{equation}
The measure becomes
\begin{equation}
S_{q}=\frac{1}{2}[V_{11}+V_{66}+V_{22}+V_{55}-V_{16}-V_{61}-V_{25}-V_{52}]^{-1},
\end{equation}
where the matrix elements of $V$ and its evolution can be derived by time integration of
its dynamical equation\cite{li2015criterion,mari2012opto,larson2011photonic,wang2014nonlinear,du2017synchronization}
\begin{equation}
\dot{V}=MV+VM^{T}+N
\label{mx}
\end{equation}
which can be directly attained from Eq.(\ref{mc}). The noise matrix $N=\mathrm{diag}(0,\gamma_{m}(2n_{b}+1),\kappa,\kappa,0,\gamma_{m}(2n_{b}+1),\kappa,\kappa)$ satisfying $N_{ij}\delta(t-t^{'})=\frac{1}{2}\langle n_{i}(t)n_{j}(t^{'})+n_{j}(t^{'})
n_{i}(t) \rangle$.

From Eq. (\ref{flu}), we can find that, unlike the linear cases\cite{PhysRevE.95.022204,PhysRevE.93.062221,du2017synchronization,geng2018enhancement}, the large nonlinearity intensity $\chi_j$ can suppress the oscillations of the two cavity as well as the photon exchange between them, and indirectly ``frozen" the oscillation of the positions and the momentums of two mechanical oscillators. In this situation, $S_q$ can directly reaches its maximal values in a very short time since the oscillations of the two mechanical oscillators are both suppressed. For small nonlinear intensities, the nonlinear terms in Eq. (\ref{flu}), (\ref{mc}) and (\ref{mx}) can also modify the evolution of $S_q$ and improve the quantum synchronization. Combine with the different modulations and couplings, we next discuss the influence of the Kerr nonlinearity on evolution of $S_q$ in more detail via the Numerical simulation of Eq. (\ref{flu}), (\ref{mc}) and (\ref{mx}).
\section*{Numerical Results and Discussion}
To examine the effects of Kerr nonlinearity, different time modulations and different couplings on the quantum synchronization, we numerically calculate the dynamics of the mean values of the fluctuations. We mainly discuss the quantum synchronization of nonlinear optomechanical system (since its classical synchronization measured by $S_c$ are nearly perfect in the following cases , it will not be presented here). Beyond the quantum linear system by periodically modulating cavity detunings or driving amplitudes \cite{farace2012enhancing,mari2012opto,liao2015enhancement,du2017synchronization}, the Kerr nonlinearity brings out some new phenomena as we adjusting the nonlinear strength $\chi_j$. The value of $\chi_j$ are restrict to small values, since the strong nonlinearity will greatly restrain the oscillation of the mechanical oscillators inspite of a perfect quantum synchronization can be expected.  Next, we will discuss the effect of Kerr nonlinearity on the quantum synchronization in different types of periodical modulation (periodically modulating cavity detunings or driving amplitudes) and different couplings (indirectly coupled mechanical oscillators through an optical fiber or directly coupled mechanical oscillators by phonon tunnel).

\begin{figure}[t]
\centering
\includegraphics[width=0.8\textwidth]{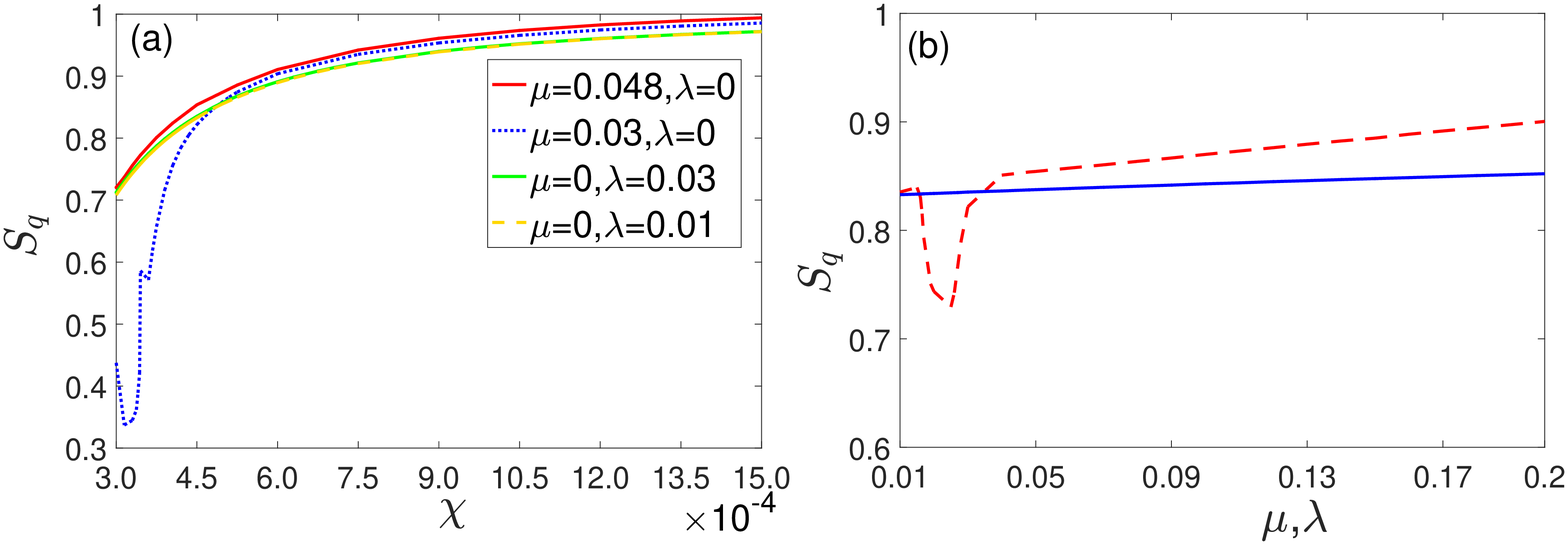}
\caption{(a) Mean values of the measure of quantum synchronization $S_{q}$
versus Kerr coupling coefficients $\chi$ with $\eta_{C}$=1, $\Omega_{C}=1$ and different coupling intensities (red solid line for $\mu=0.048$, $\lambda=0$, the blue dotted line for $\mu=0.03$, $\lambda=0$, green solid line for $\mu=0$, $\lambda=0.03$, yellow dashed line for  $\mu=0$,$\lambda=0.01$)
 (b) Mean values of $S_{q}$
versus the phonon tunneling intensity $\mu$ (red dashed line) and the
coupling constant of cavity modes $\lambda$ (blue solid line) with $\chi=0.00045$.
Other parameters are chosen as $\triangle_{1}=1$,$\triangle_{2}=1.005$ ,$\omega_{j}=\triangle_{j}$, $g1=g2=0.005$, $E=100$.}
\label{2}
\end{figure}

\subsection*{Modulation on cavity detunings. ($\eta_{D}=0$, $\eta_{C}\neq0$)}
We first consider the case of modulating the cavity detunings ($\eta_{C}=0.5$, $\Omega_{C}=1$) and leave the driving fields unchanged. For simplicity, $\chi_1$ and $\chi_2$ are assumed to be equal, i.e. $\chi_1=\chi_2=\chi$, and $\omega_1=\Delta_1$ and $\omega_2=\Delta_2$ can be slightly different.

\begin{figure}[t]
\centering
\includegraphics[width=0.8\textwidth]{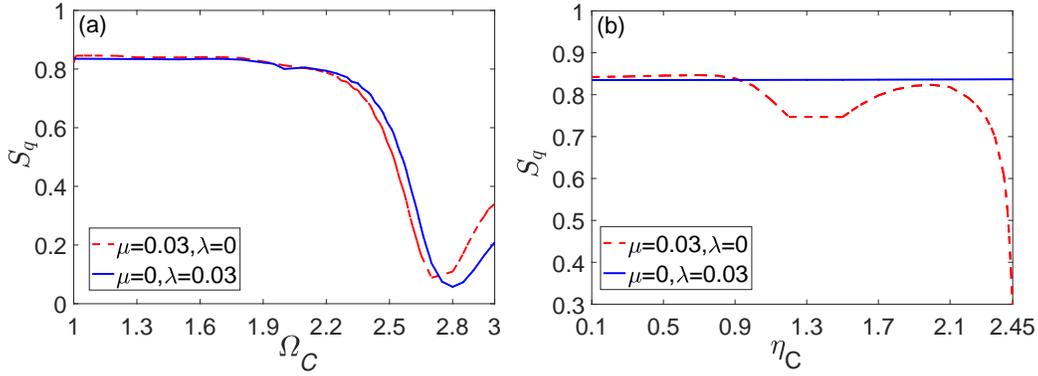}
\caption{(a) Mean values of quantum synchronization $S_{q}$ measures versus modulation frequency $\Omega_{C}$ with $\eta_{C}=1$. (b) Mean values of quantum synchronization $S_{q}$ measures modulation amplitude $\eta_{C}$ with $\Omega_{C}=1.0$. The other parameters are
the same as in Fig.\ref{2}.
}
\label{3}
\end{figure}
\begin{figure}[t]
\centering
\includegraphics*[width=0.7\columnwidth]{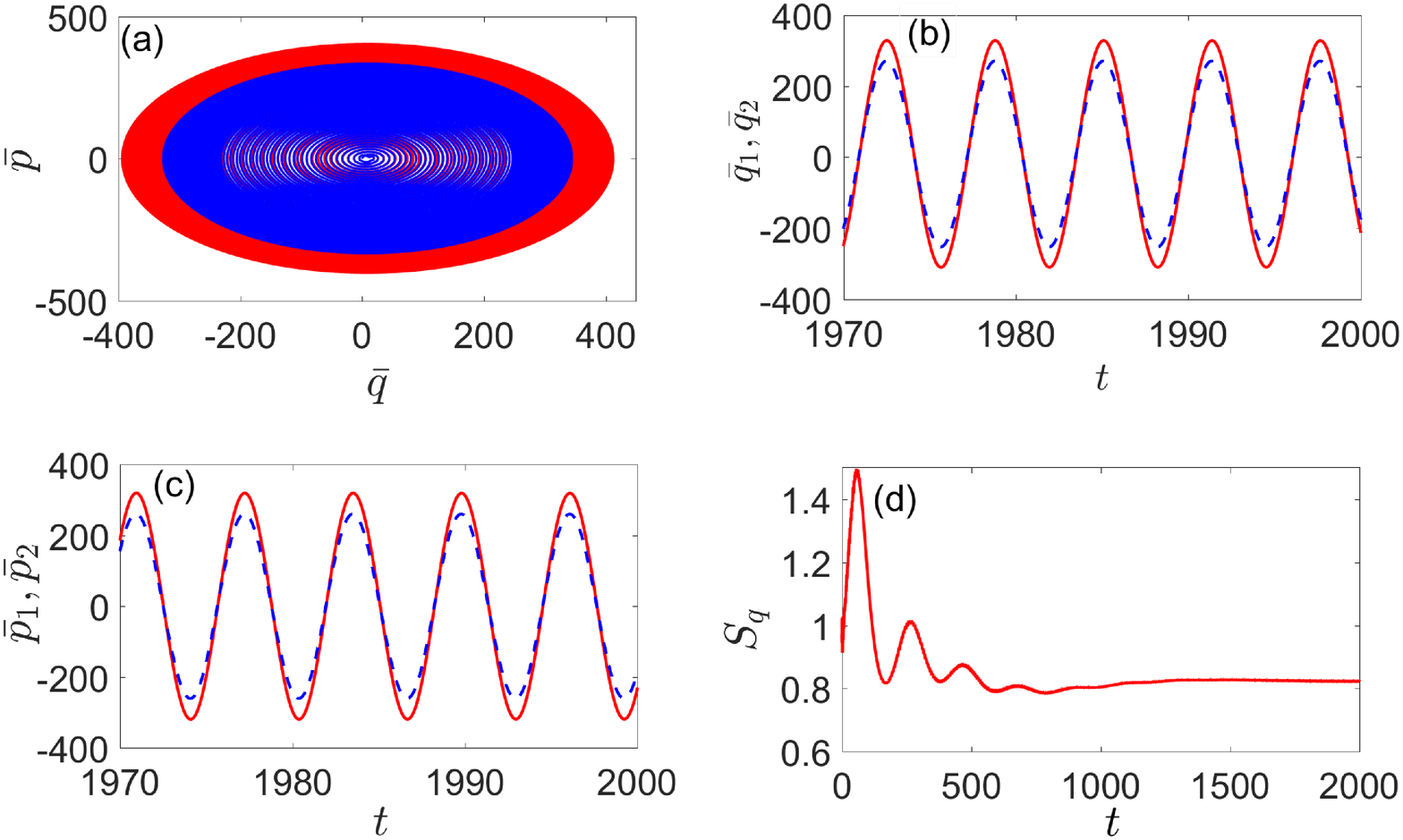}
\caption{(a) The evolution of the mean values $\bar{q}(t)$ and $\bar{p}(t)$ of the
two mechanical oscillators position and momentum (blue and red lines) with directly coupling. (b) Time evolution of the mean
value $\bar{q}_{1}(t)$(red solid line) and $\bar{q}_{2}(t)$(blue dashed line). (c) Time evolution of the mean value $\bar{p}_{1}(t)$(red solid line) and $\bar{p}_{2}(t)$(blue dashed line). (d)
Time evolution of $S_{q}(t)$. Here we set $\Omega_{C}=1, \eta_{C}=1, \mu=0.03, \lambda=0$ and the other parameters are
the same as in Fig.2
}\label{4}
\end{figure}

\begin{figure}[!hbt]
\centering
\includegraphics*[width=0.7\columnwidth]{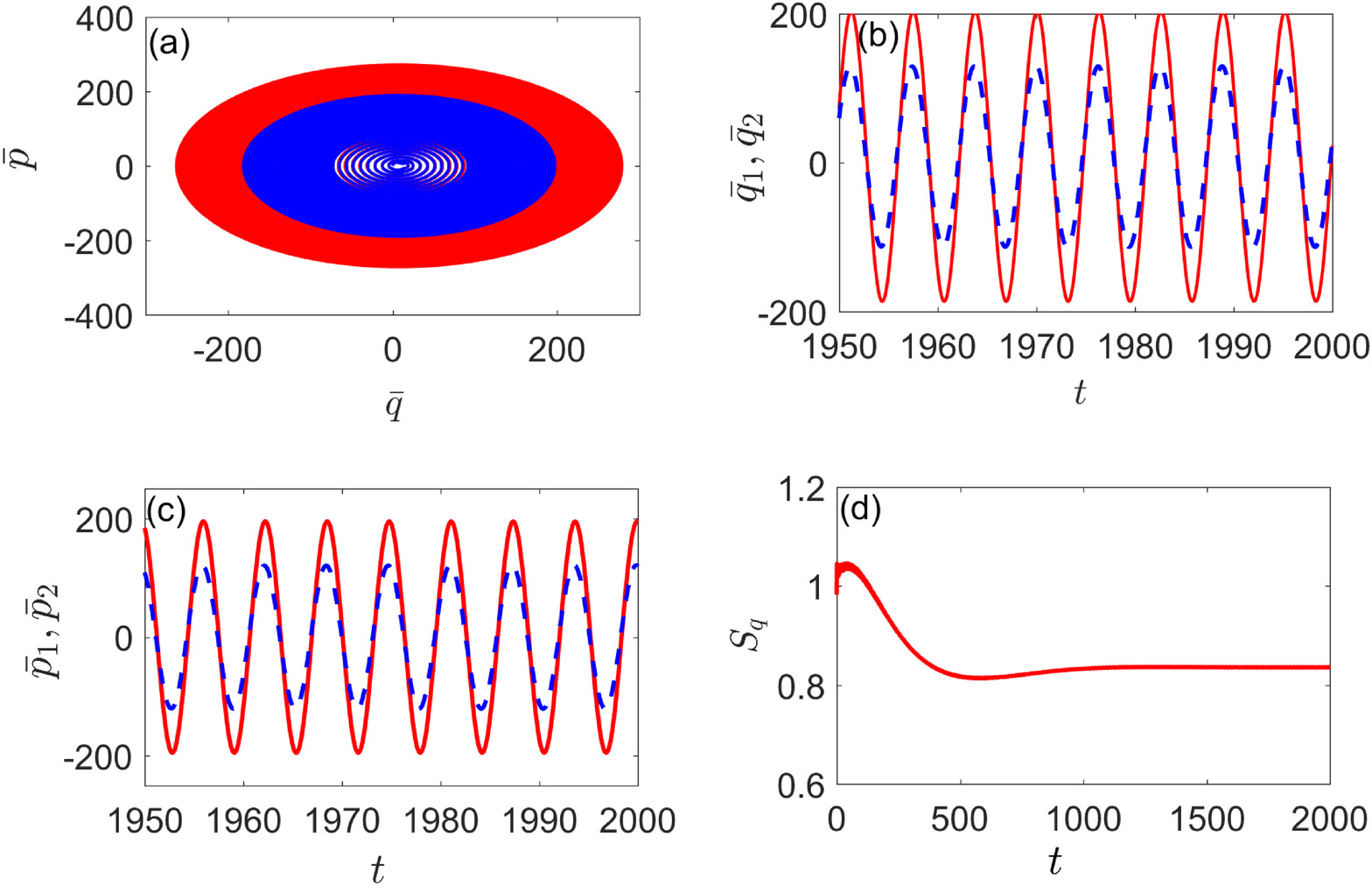}
\caption{(a) The evolution of the mean values $\bar{q}(t)$ and $\bar{p}(t)$ of the
two mechanical oscillators position and momentum (blue and red lines) with indirectly coupling. (b) Time evolution of the mean
value $\bar{q}_{1}(t)$(red solid line)  and $\bar{q}_{2}(t)$(blue dashed line). (c) Time evolution of the mean value $\bar{p}_{1}(t)$(red solid line)  and $\bar{p}_{2}(t)$(blue dashed line). (d)
Time evolution of $S_{q}(t)$. Here we set $\Omega_{C}=1, \eta_{C}=1, \mu=0, \lambda=0.03$ and the other parameters are
the same as in Fig.2
}\label{5}
\end{figure}

As shown in Fig. \ref{2}(a), no matter how the two optomechanical system are coupled (directly or indirectly), the quantum synchronization can be continually enhanced by the increasing of the nonlinear intensity $\chi$. Moreover, for direct coupling and $\chi$ less than $0.00045$, the directly coupling coefficients $\mu$ can significantly influence the degree of quantum synchronization. In contrast, the indirect coupling coefficients $\lambda$ is more stable for the quantum synchronization. For $\chi\ge0.00045$, $S_{q}$ can reach $0.8$ or larger for any kind of coupling (directly or indirectly). This proves that the Kerr nonlinearity can significant improve the quantum synchronization.
Note that the dependence of $S_q$ on $\chi(>0.00045)$ with different coupling types and intensities are similar,  we choose $\chi=0.00045$ to study the dependence of $S_{q}$ on the directly coupling coefficient $\mu$ and indirectly coupling coefficient $\lambda$ [see Fig. \ref{2}(b)]. It is easy to be found that the
quantum synchronization can be slightly enhanced for appropriate values of $\mu$ and $\lambda$ ($\mu$ is more significant). The same coupling intensity exist some differences for different coupling types as the coupling coefficient changes. However, both of them have good quantum synchronization effects ($S_{q}>0.8$).

Besides, the modulation frequency $\Omega_{C}$ and modulation amplitude $\eta_{C}$ also play important roles in quantum synchronization. As shown in Fig. \ref{3}, the quantum synchronization is better for small $\eta_{C}$ ($1 \sim 2$) and $\Omega_{C}$ ($0.1 \sim 0.9$). While it becomes unstable and worse for greater $\eta_{C}$ ($2 \sim 3$) and $\Omega_{C}$ ($0.9 \sim 2.45$). For the same coupling intensity, indirect coupling is better than direct coupling for quantum synchronization. This means that suitable modulations on cavity detunings are also needed for a good quantum synchronization. Since the nonlinearity brought by the Kerr medium is unstable, the values of the quantum synchronization with larger external disturbances will become chaotic.
For small amplitude and frequency modulation, $S_{q
}$ can easily reaches $0.85$ for both of the two coupling ways, which is better than the optimal values of the corresponding linear systems that are modulated \cite{du2017synchronization,geng2018enhancement}. Of course, we can continue to increase the nonlinear intensity to further improve quantum synchronization [see Fig.\ref{2}(a)]. However, a too-large value of nonlinear strength will suppress the oscillating amplitudes of the two mechanical oscillators which is clearly not what we wanted. In practical applications, the perfect synchronization need its degree greater than $0.9$. After some numerical simulations, we find that $\chi=0.0006$ can be chosen as an optimal value of the nonlinearity intensity ($S_{q}>0.9$, amplitudes of the oscillations of $\bar{q}$ and $\bar{p}$  are both greater than 100).

In order to investigate the dynamics of the system in synchronization, we further examine the evolution of the mean values position($\bar{q}$) and momentum($\bar{p}$) of the two oscillators with $\Omega_{C}=1$ and $\eta_{C}=1$. The mean values position $\bar{q}_{1}(t)$ and $\bar{q}_{2}(t)$ as well as the mean values momentum $\bar{p}_{1}(t)$ and $\bar{p}_{2}(t)$ are found to be oscillating with exactly the same phases in the stable state as shown in Fig. \ref{3}(b) [Fig. \ref{4}(b)] and Fig. \ref{3}(c) [Fig. \ref{4}(c)]. Meanwhile, two corresponding limit-cycle trajectories of the two mechanical oscillators in phase space are illustrated in the inset of Fig. \ref{4}(a) [Fig. \ref{5}(a)]. As shown in Fig. \ref{4}(d) [Fig. \ref{5}(d)], the system will reaches a steady state in the end and $S_{q}$ tends to a stable value. (the initial covariance matrix is randomly generated and unnormalized since we are only interested in the steady state). This means that, with the existence of Kerr nonlinearity, the degree of quantum synchronization between two mechanical oscillators with different frequencies can also be enhanced by periodically modulating cavity detunings with appropriate parameters.

\begin{figure}[t]
\centering
\includegraphics[width=0.8\textwidth]{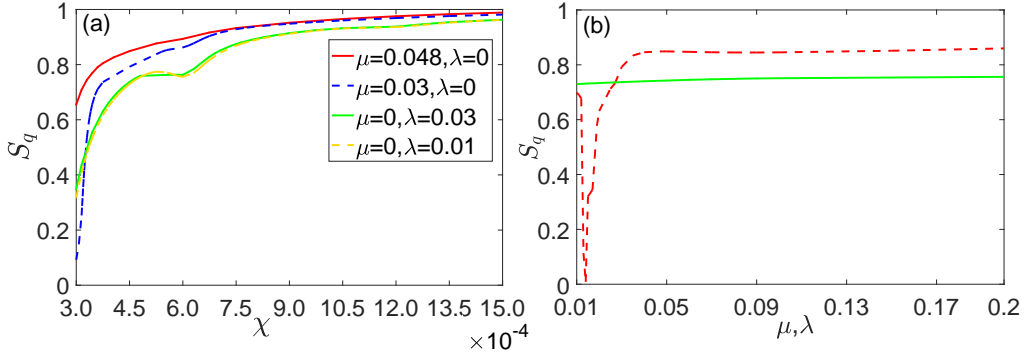}
\caption{(a) Mean values of quantum synchronization $S_{q}$ measures
versus second-order nonlinear optical detunings $\chi$ with $\eta_{D}$=0.5, $\Omega_{D}=1$ and different coupling (The red line $\mu=0.048$, $\lambda=0$,the blue line $\mu=0.03$, $\lambda=0$, the green line $\mu=0$, $\lambda=0.03$, the yellow line $\mu=0$, $\lambda=0.01$) (b) Mean values of quantum synchronization $S_{q}$ measures
versus  a phonon tunneling term of intensity $\mu$ (red dashed line) and the
coupling constant of cavity modes $\lambda$ (green solid line).
Some parameters are $\triangle_{1}=1$,$\triangle_{2}=1.005$ ,$\omega_{j}=\triangle_{j}$, $g1=g2=0.005$, $E=100$.}
\label{6}
\end{figure}

\begin{figure}[t]
\centering
\includegraphics[width=0.8\textwidth]{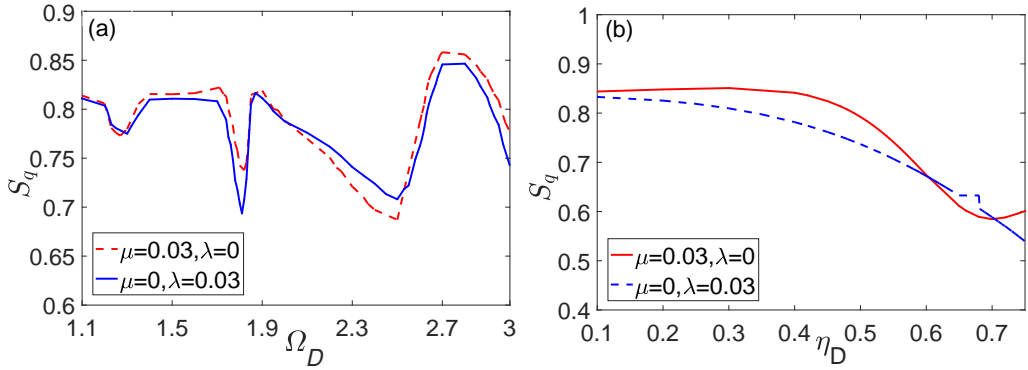}
\caption{(a) Mean values of quantum synchronization $S_{q}$ measures versus modulation frequency $\Omega_{D}$ with $\eta_{D}=0.5$. (The red line $\mu=0.03$, $\lambda=0$, the blue line $\mu=0$, $\lambda=0.03$). (b) Mean values of quantum synchronization $S_{q}$ measures modulation amplitude $\eta_{D}$ with $\Omega_{D}=1.0$. (The red line $\mu=0.03$, $\lambda=0$, the blue line $\mu=0$, $\lambda=0.03$) and the other parameters are
the same as in Fig.6
}
\label{7}
\end{figure}

\subsection*{Modulation on driving amplitudes. ($\eta_{C}=0$, $\eta_{D}\neq0$)}
Alternatively, we can periodically modulate the amplitudes of driving fields to investigate the nonlinear effect on quantum synchronization. As shown in Fig. \ref{6}(a), similar with the case of cavity detuning modulations, the quantum synchronization can be continually enhanced with the increase of nonlinear intensity $\chi$, its degree can be influenced significantly by the directly coupling coefficients $\mu$ with $\chi\le 0.00045$, and the indirect coupling $\lambda$ has little effect on the quantum synchronization in a certain nonlinear intensity. While, different with the case of cavity detuning modulations, $S_q$ can reach a steady value more rapidly as $\chi$ increases under the directly coupling, and the quantum synchronization is always better under direct than indirect coupling for the same coupling strength in most of the range $(3.0\sim4.5)$ of $\chi$. For the fixed nonlinear strength $\chi=0.00045$, quantum synchronization is more sensitive for the direct coupling $\mu$ when it goes from $0.01$ to $0.05$, For larger $\mu$, there is no significant change in the quantum synchronization of the system [see Fig. \ref{6}(b)].  Besides, unlike the stability of the quantum synchronization on the field frequency and amplitude under cavity detuning modulations, the quantum synchronization will fluctuate within a range ($0.68 \sim 0.86$) as the modulation frequency increasing [see Fig. \ref{7}(a)], and the nonlinear system is more sensitive to the modulation intensity of the field . When the amplitude of the modulation field is not great, the quantum synchronization effect is also better for the two types of coupling [see Fig. \ref{7}(b)]. Therefore, we set $\Omega_{D}=2.7$, $\eta_{D}=0.5$, the corresponding optimal values are $S_{q}\simeq 0.86$. A large driving field strength will destroy the quantum synchronization of the system.

\begin{figure}[t]
\centering
\includegraphics*[width=0.7\columnwidth]{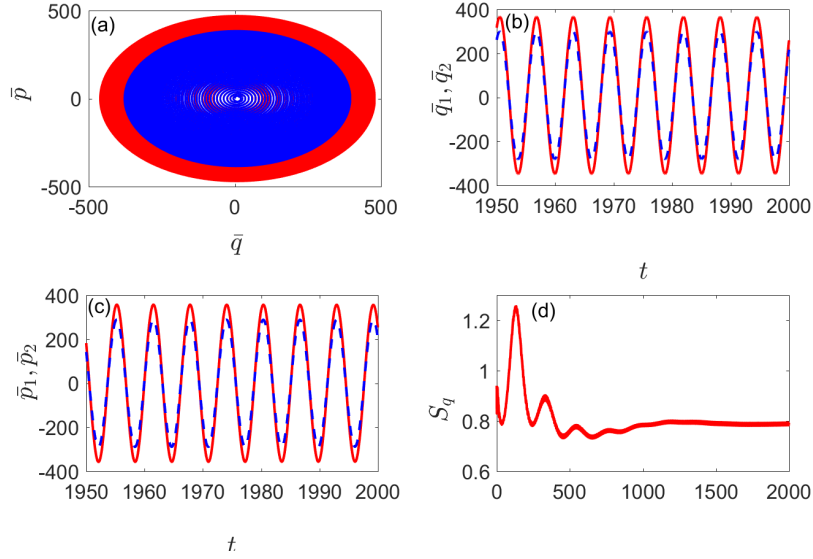}
\caption{(a) The evolution of the mean values $\bar{q}(t)$ and $\bar{p}(t)$ of the
two mechanical oscillators position and momentum (blue and red lines). (b) Time evolution of the mean
value $\bar{p}_{1}(t)$(red solid line)  and $\bar{p}_{2}(t)$(blue dashed line). (c) Time evolution of the mean value $\bar{q}_{1}(t)$(red solid line)  and $\bar{q}_{2}(t)$(blue dashed line). (d)
Time evolution of $S_{q}(t)$. Here we set $\Omega_{D}=1, \eta_{D}=0.5, \mu=0.03, \lambda=0$ and the other parameters are
the same as in Fig.6
}\label{8}
\end{figure}

\begin{figure}[t]
\centering
\includegraphics*[width=0.7\columnwidth]{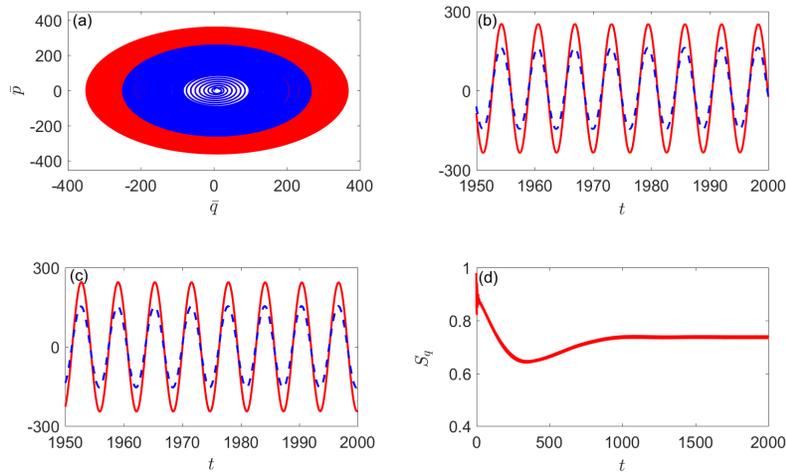}
\caption{  (a) The evolution of the mean values $\bar{q}(t)$ and $\bar{p}(t)$ of the
two mechanical oscillators position and momentum (blue and red lines). (b) Time evolution of the mean
value $\bar{q}_{1}(t)$(red solid line)  and $\bar{q}_{2}(t)$(blue dashed line). (c) Time evolution of the mean value $\bar{p}_{1}(t)$(red solid line)  and $\bar{p}_{2}(t)$(blue dashed line). (d)
Time evolution of $S_{q}(t)$. Here we set $\Omega_{D}=1, \eta_{D}=0.5, \mu=0, \lambda=0.03$ and the other parameters are
the same as in Fig.6
}\label{9}
\end{figure}

The degrees of different coupling ways under modulation of driving fields can be intuitively shown by the dynamics of mean values position and momentum of each mechanical oscillator. we set $\Omega_{C}=1$ and $\eta_{C}=1$ to compare with the situation under the modulation of cavity detuning.  As shown in Fig. \ref{8}(b) [Fig. \ref{9}(b)]
and Fig. \ref{8}(c) [Fig. \ref{9}(c)], when the system is stable, $\bar{p}_{1}$ and $\bar{p}_{2}$ are the same phase, but the amplitude is different. And $\bar{p}$ and $\bar{q}$ have similar variations
from Fig. \ref{8}(a) [Fig. \ref{9}(a)], the evolution of phase diagram is two limit-cycle trajectories, which are slight difference and from Fig. \ref{8}(d) [Fig. \ref{9}(d)], we can see that the system reaches a
steady state in the end and $S_{q}$ tends to a stable value over time. It is easy to find that the degree of quantum synchronization is better under direct than indirect coupling with the same nonlinear strength, modulation frequency and amplitude. Nevertheless, the nonlinearity and the periodical modulation on driving field can always enhance the quantum synchronization.

\subsection*{Comparison of two modulations}
Now let's compare the types of quantum synchronization in the nonlinear optomechanical system with the two different ways of periodical driving. Comparing Fig. \ref{3} and Fig. \ref{7}, we find that small amplitude or frequency of the periodic modulation has a better effect on quantum synchronization and the system is more stable under a certain nonlinear intensity. However, when the amplitude or frequency of the periodic modulation is large, quantum synchronization has different changes in the two different modulation. Simultaneously, quantum synchronization has a slight enhancement (a large change) through indirect coupling (direct coupling) as the coupling coefficient $\lambda$ ($\mu$) increases and when the nonlinear intensity $\chi$ exceeds a certain value, the quantum synchronization is not affected by the coupling coefficient $\mu$ or $\lambda$ (see Fig. \ref{2} and Fig. \ref{6}). According to the above analysis, we find that dynamics of the nonlinear system is correspondingly more sensitive to the change of the modulation of driving fields amplitude and the direct coupling mode.

\section*{Conclusions}
In summary, we have studied the quantum synchronization phenomenon of mechanical oscillators of different
frequencies in nonlinear optomechanical system by periodically modulating the cavity detunings or the driving field in two different ways of coupling.
After detailed analysis and comparing to the former studies \cite{du2017synchronization,geng2018enhancement}, we find that the coupled optomechanical systems with Kerr nonlinearity under appropriate modulations
on cavity detunings or driving amplitudes has better degrees of quantum synchronization than the linear one, and it is also easier to enhance the quantum synchronization effect and realize good quantum synchronization effect ($S_{q}>0.8$) for two different ways of coupling (direct coupling and indirect coupling). Nevertheless, the direct coupling $\mu$ and indirect coupling $\lambda$ coefficient have different effects on quantum synchronization: the former haves a large adjustment range and the latter is more stable. The two different modulation ways can also lead to different behaviors of quantum synchronization with the same parameters. The dynamics of the system is more sensitive to the modulation of driving fields amplitude. In any way, the quantum synchronization can be improved by increasing the nonlinear intensity and the value of $S_{q}$ can be approximated to 1. Therefore, we believe that the study of Kerr nonlinearity and its effect on the quantum synchronization  may have a further promoting effect on quantum communication and quantum control.


\section*{Acknowledgements}
This work is supported by National Natural Science Foundation of China (NSFC) (Grants No. 11875103, No. 11775048, and No. 61475033) the Plan for Scientific and Technological Development of Jilin Province (Grant No. 20160520173JH), and the Scientific and Technological Program of Jilin Educational Committee during the Thirteenth Five-year Plan Period (Grant No. JJKH20180009KJ).

\section*{Author Contributions}
H.-D. Liu and X. X. Yi initiated the idea. G.-J. Qiao and H.-D. Liu wrote the main manuscript text. G.-J. Qiao and H.-X. Gao performed the calculation.

\section*{Additional Information}
\textbf{Competing Interests:} The authors declare that they have no competing interests.

\end{document}